# Few-photon imaging at 1550 nm using a low-timing-jitter superconducting nanowire single-photon detector

**Hui Zhou, Yuhao He, Lixing You,** *** Sijin Chen, Weijun Zhang, Junjie Wu, Zhen Wang, and Xiaoming Xie**

*State Key Laboratory of Functional Materials for Informatics, Shanghai Institute of Microsystem and Information Technology, Chinese Academy of Sciences, 865 Changning Road, Shanghai 200050, China*
*\*lxyou@mail.sim.ac.cn*

**Abstract:** We demonstrated a laser depth imaging system based on the time-correlated single-photon counting technique, which was incorporated with a low-jitter superconducting nanowire single-photon detector (SNSPD), operated at the wavelength of 1550 nm. A sub-picosecond time-bin width was chosen for photon counting, resulting in a discrete noise of less than one/two counts for each time bin under indoor/outdoor daylight conditions, with a collection time of 50 ms. Because of the low-jitter SNSPD, the target signal histogram was significantly distinguishable, even for a fairly low retro-reflected photon flux. The depth information was determined directly by the highest bin counts, instead of using any data fitting combined with complex algorithms. Millimeter resolution depth imaging of a low-signature object was obtained, and more accurate data than that produced by the traditional Gaussian fitting method was generated. Combined with the intensity of the return photons, three-dimensional reconstruction overlaid with reflectivity data was realized.

## 1. Introduction

Light detection and ranging (LIDAR) is a remote sensing technology that measures distance by illuminating a target with a laser and analyzing the time-of-flight (TOF) of the reflected light. This method is used to make high-resolution maps for many applications, including remote sensing, terrain mapping, and space debris tracking [1, 2]. Combined with advanced time-correlated single-photon counting (TCSPC) techniques, LIDAR is sensitive to ultra-low levels of light and exhibits excellent depth resolution, which are determined by the detection efficiency (DE) and the timing jitter (TJ) of the optical receivers. Laser ranging [3-5] and three-dimensional (3D) imaging [6-9] based on TCSPC have been progressively investigated in recent years, exhibiting shot-noise limited detection and good surface-to-surface resolution on the millimeter scale.

Most LIDAR applications with outstanding performance have been realized at wavelengths under 1000 nm using a Si avalanche photodiode (APD) [3, 10, 11] or a photomultiplier [12, 13]. Recent progress in erbium-doped lasers suggests burgeoning various applications at the wavelength of 1550 nm, which lies in the eye-safe region of the spectrum and offers much lower solar background noise, as well as lower atmospheric attenuation. Investigations on TCSPC range profiling [14] and depth imaging [6] at this wavelength, in which a single InGaAs/InP APD or APDs in a focal plane array architecture [15, 16] were often applied, have been carried out in recent years. However, InGaAs/InP APDs, even when operated in gated Geiger mode to maintain better performance, often suffer from a large dark count rate (DCR), afterpulsing effects, and poor timing resolution, resulting in a relatively low surface-to-surface resolution.

In the last decade, superconducting nanowire single-photon detectors (SNSPDs) have attracted considerable attention because of their high DEs, low DCRs, no afterpulsing, and low-TJ. SNSPDs are considered as a promising alternative to conventional single-photon detectors (SPDs) and have been successfully used in near-infrared laser ranging [5] and depth imaging [7]. Nevertheless, infrared depth imaging based on the TCSPC technique usually required hundreds of detected photons per pixel to obtain accurate information about the range and reflectivity, as the reported imaging system used a relatively noisy SPD with a high TJ in the receiver. The recently reported depth imaging at 1560 nm used an SNSPD with a TJ of 98 ps and a DCR of 1 kHz and obtained centimeter resolution depth images of low-signature objects in daylight at stand-off distances of the order of one kilometer [7]. Pixel-by-pixel information was obtained by fitting the histograms of time delays between the transmitted and detected photons, and the surface-to-surface resolution was directly determined by the system TJ. Tens of signal photons per pixel were the minimal requirement for a correct fitting with reasonable background noise counts. In cases of the low-signature target return detection with a relatively high TJ detector, specific algorithms, such as the cross-correlation algorithm [9, 10], had to be employed to analyze the presence of a signal in a high background level so as to improve the accuracy of the measurement.

In general, the depth resolution of an imager is predominantly determined by the TJ of the detector. SPDs with relatively low-TJs may not only provide a higher depth resolution but also improve the signal-to-noise ratio. Indeed, the noise count per time bin, contributed by the stray light, has a random feature and can be decreased by employing a smaller time bin. In extreme cases, the target signal can be determined if the noise count per time bin is no more than 1, and the signal count is greater than 2. In this way, no fitting or algorithm is necessary. However, a low-TJ system is a prerequisite for realizing this simple yet effective method.

In this paper, we demonstrated a TCSPC depth imager operated at a wavelength of 1550 nm that incorporated a low-TJ SNSPD. The TJ of the optimized system was 44 ps full width at half maximum (FWHM). An 813-fs time-bin width was chosen for photon counting, which gave discrete noise of less than one count in indoor daylight condition and less than two counts in outdoor daylight condition for a 50-ms collection time. Because of the low-TJ SNSPD and sub-picosecond time bin, the depth information was determined directly by the highest bin counts, which produced more accurate data than the traditional Gaussian fitting method for the low-signature target. Combined with the intensities of the return photons, the indoor and outdoor depth imaging overlaid with reflectivity data for a target with a stand-off distance of 2.5 m were demonstrated.

## 2. Experimental setup and results

The experiment setup is schematically illustrated in Fig. 1. A 20-MHz mode-locked fiber laser (FPL-01CAF, Calmar) centered at 1550 nm, with a pulse duration of 500 fs, and a typical maximum average output power of 2 mW was employed as the illumination source in our laser depth imaging system. The output beam was expanded and collimated with a single-mode fiber (SMF, 9-μm fiber core) pigtailed reflective collimator with an output beam size of 6 mm. The optical transceiver was coaxial and consisted of two reflective mirrors (M1 and M2). We used a commercial polyvinyl chloride (PVC) panda as the target object, which was placed at a distance of 2.5 m from M2. The beam size on the target was considered to be unchanged at this target distance. The size of the target was $15 \times 10 \times 22$ cm (length × width × height, viewed from the side), denoting that depth profiling could be achieved without the concern of range ambiguity, which was limited to 7.5 m by the 20-MHz laser frequency. To realize the depth imaging of the target, the target was placed on the two-dimensional platform driven by stepper motors, which was controlled by the computer to complete the X-Y scan. The spatial resolution determined by the stepper motor was 25 μm. The scattered target return photons were collected with a reflection Newton telescope (150-mm aperture diameter), and filtered by an optical bandpass interference filter (FWHM ~ 12 nm) centered at 1550 nm to block the ambient light. The collected photon flux was then converged by a lens (3-cm focal length) and coupled into a multi-mode fiber (MMF), with a core size of 50 μm, which was connected to an SMF because the SNSPD was optically coupled with the SMF. The total optical loss of the system was about 30 dB, including the 20-dB coupling loss from MMF to SMF. The output of the SNSPD was connected to the "stop" of the TCSPC module, while the synchronous trigger signal of the laser source was linked to the "start" of TCSPC. All the measurements reported in this paper were taken while operating with an average emitted optical power of about 100 μW.

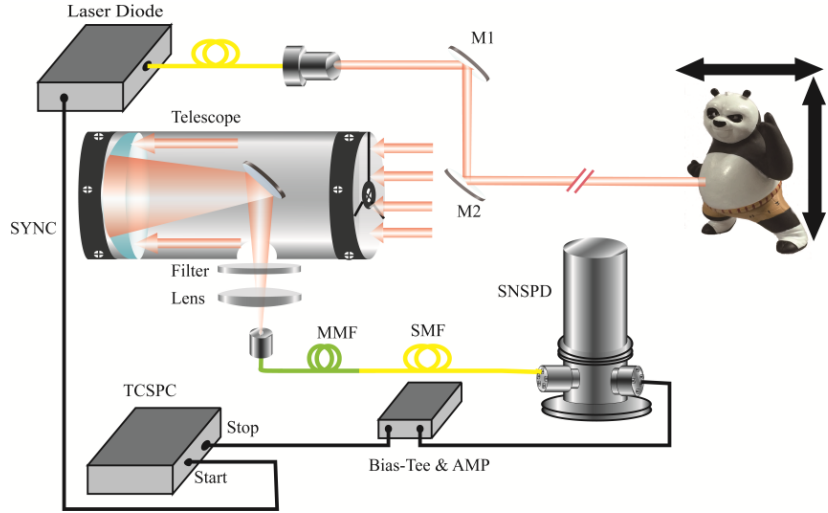

Fig. 1. Schematic of 1550 nm laser depth imaging system based on a TCSPC module using SNSPD. M1 and M2, high-reflectivity gold mirrors with 25-mm diameter; MMF, mutimode fiber; SMF, single-mode fiber; SYNC, synchronous trigger signal of the laser source.

The SNSPD employed in the TCSPC system was a 90-nm wide meander-structured nanowire made of a 7-nm-thick NbN film, covering an active area of $15 \times 15$ μm$^2$, with a filling factor of 37.5%. The SNSPD was coupled with SMF and mounted inside a closed-cycle Gifford–McMahon cryocooler system, which was operated at 2.21 ±0.02 K. The system detection efficiency (SDE) of SNSPD at 1550 nm was 29%, with a DCR of 20 Hz, when the detector was biased at 90% of its switching current (20.3 μA). The output voltage pulses from SNSPD were amplified by a low-noise amplifier (LNA650 from RF Bays Inc: 30 k–600 MHz, G=50 dB) at room temperature, before being routed to the "stop" port of the TCSPC module.

To lower the TJ of the imaging system, we chose SPC-150 (Becker & Hickl GmbH) as the data acquisition module. This module had a lower TJ of 7.6-ps FWHM and was configured with a minimal time bin of 813 fs, as opposed to the popular TCSPC module PicoHarp 300 (PicoQuant GmbH), which had a TJ of 26-ps FWHM and a 4-ps time bin per channel. In addition, the TJs of the employed femtosecond-pulsed laser and its synchronization trigger were 60 fs and 4 ps, respectively.

The retro-reflected photons were detected by the SNSPD, and the corresponding histograms were then recorded by the SPC-150 over a fixed integration time (with a time bin per channel of 813 fs). Figure 2 gives an example of the normalized statistical distribution, which indicated a system TJ of 44 ps. Although the SNSPD had a TJ higher than the value reported in our previous laser ranging system [5, 17], its SDE was higher by one order of magnitude. Besides, the TJ was smaller compared with the values reported in other studies [2, 7]. Such a low-TJ TCSPC system could effectively reduce the uncertainty of the 3D reconstruction.

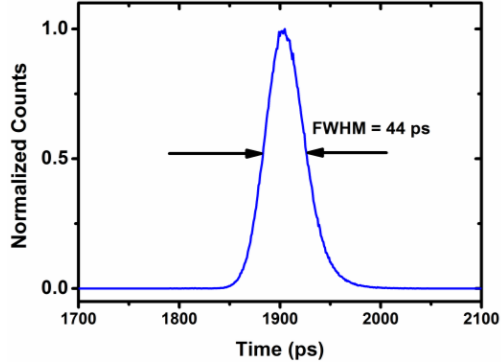
Fig. 2. Normalized instrumental response of the system with a TJ of 44-ps (FWHM).

In order to give an intuitive result of the system depth resolution, two retro-reflected mirrors were adopted to replace the target for the measurement. We adjusted the separations of the two mirrors, and the typical photon-return histograms were recorded by TCSPC and shown in Fig. 3. The red dashed lines and blue solid lines corresponded to the two-peak Gaussian fittings of the time correlation for the two mirrors. By locating the centroids of each of the two return signals, the separations between them were obtained. The distance was calculated using the formula $L = c \times \Delta t / 2$, where $c$ is the speed of light in air and $\Delta t$ is the time interval of the two centroids of the return signals. When the separation was 30 mm, the two peaks was clearly separated by a time interval of 200 ps, as shown in Fig. 3(a). For the separation of 6.6 mm (Fig. 3(b)), corresponding to 44 ps time interval which was also the system FWHM jitter, the two peaks were still easy to distinguish from one another. When the separation was further reduced to 3 mm, even the two peaks overlapped, we could differentiate them using the two-peak Gaussian fittings, as shown in Fig. 3(c). Thus, we may conclude that the depth resolution was about 5 mm. Optimized signal-processing algorithm based on a reversible-jump Markov chain Monte Carlo method may provide a better depth resolution [2].

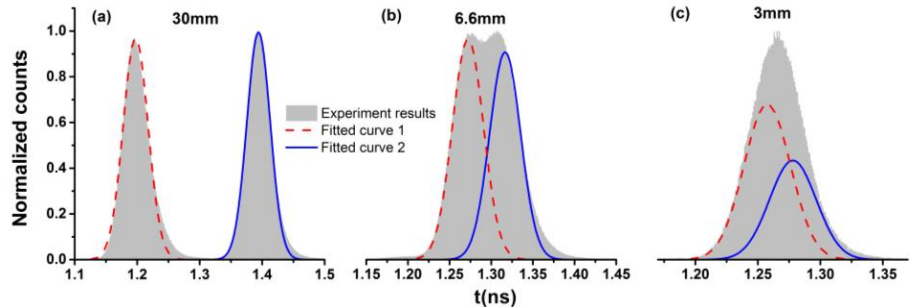
Fig. 3 Time correlation results from the targets of different surface separations.

The depth information was obtained for each individual pixel while the target was scanned in the X-Y plane. A depth profile of the scanned target could then be reconstructed on a computer. Besides, the pixel-by-pixel relative reflectivity could be determined from the number of photons detected in a fixed dwell time. The number of collected photons varied with the target surface material, roughness, and incidence angle. Figure 4(a) shows a comparison of the return photon counts in 44-ps time bins, with logarithmic coordinates at the different positions indicated in Fig. 4(b). Generally speaking, the counts varied by over three orders of magnitude; for example, the return photon count from the background paper (Location A) was 200 times larger than that from the center on the black neck part of the

panda(Location B). The different color caused intensity variance can be found in location B&D. The different incident angle caused intensity variance can be found in location C&D. At the edge, with the surface normal nearly perpendicular to the incidence of the light, the difference was even larger. Though locations A&D have the similar white color, the reflected photon counts has a large difference due to the different reflectivity, i.e. different materials. The main challenge involved in the depth profiling was how to get depth information from a low-signal count when it was difficult to distinguish from the noise count. Some algorithms were introduced to improve SNR and minimize the depth uncertainty when these systems were applied in a high background level and were relatively noisy with a TJ on the order of hundreds of picoseconds [9, 18]. Here, we showed how to realize laser depth imaging with high accuracy using low-jitter SNSPD directly without any fitting method and/or algorithms.

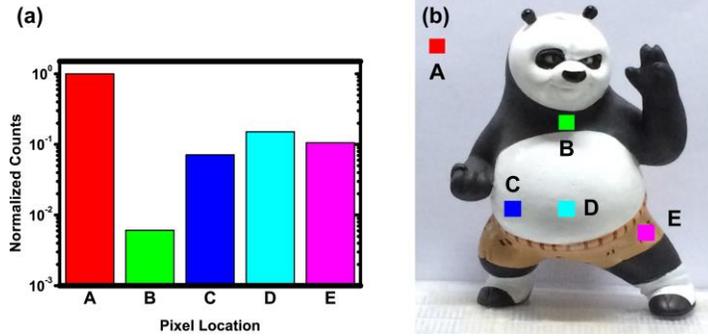

Fig. 4. (a) Integrated photon numbers with logarithmic coordinates at different locations of the target, and (b) the corresponding locations on the imaging target. The optical image of the PVC panda was taken by the first author (H ZHOU) using an optical camera.

To improve the system, a low-TJ SNSPD was used in combination with the 813-fs time bin. The sub-picosecond time bin provided discrete counts of no more than 1 for each time bin at a dwell time of 50 ms under daylight indoor conditions. Because of the 44-ps TJ of this system, we could obtain enough signal counts using the sub-picosecond time bin at a dwell time of 50 ms, which could be discriminated from the noise count easily as long as it was over 2. Figure 5(a) shows a typical example of a few signal photons with a collection time of 50 ms. The sum of the stray-light photons and intrinsic detector dark counts was about 300 Hz. For a collection time of 50 ms, only 15 noise counts were randomly distributed throughout the 50-ns pulse interval. Since we chose 3.3 ns as the observation window, the average number of noise counts was only 1 in an average of 4096 time bins. As a result, it was reasonable for us to observe only 1 noise count in a 3.3-ns observation window (Fig. 5(a)). In fact, no noise counts were observed for most of time during the measurement. Consequently, a very weak target signal could be easily distinguished from the noise, provided that the maximal count in one time bin was greater than or equal to 2 ($\geq$2); thus, a sufficient accuracy for depth profiling was achieved because of the low-TJ of 44 ps. Indeed, this method is also applicable for outdoor daylight conditions. Figure 5(b) shows the stray light noise counts (roughly ~75 KHz) recorded under outdoor daylight conditions (in a sunny day), with a collection time of 50 ms, using the same time bin. The probabilities of seeing 2, 1, and 0 noise photons in each time bin were 0.17%, 5.75%, and 94.08%, respectively, indicating that if we set the threshold higher than 4, or even 3, we may easily obtain the correct depth information. The outdoor noise counts in cloudy weather and evening were also evaluated, which were about 30 KHz and 400 Hz, respectively.

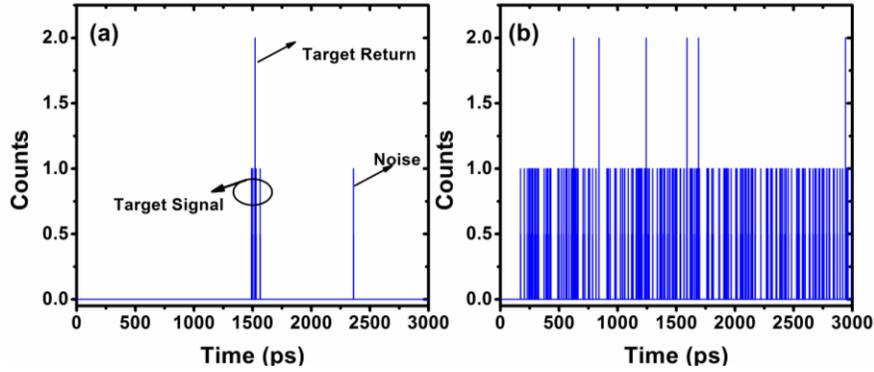

Fig. 5. (a) The weakest signal counts distinguishable in the system in the 3.3-ns observation window with a 50-ms collection time. (b) Background noise counts recorded by the TCSPC system in outdoor daylight with a collection time of 50 ms.

Following the above simple discrimination method, we could realize laser depth imaging without the use of fitting and/or algorithms. A 3D depth profile for the target is shown in Fig. 6(a). The depth scan was acquired in a room with an incandescent lamp as the background light. The scan covered an area of 200 mm × 220 mm using 200 × 220 pixels, resulting in a pixel-to-pixel spacing of 1 mm in x and y. The dwell time was set to 50 ms per pixel to accomplish a quick scan. About 99.9% of the 200 × 220 pixels exhibited reasonable depth measurements with our discrimination method, when the discrimination threshold was set to 2. For the left 0.1% pixels with the maximal count of 1 in one time bin, we averaged the arrival time of all the time bins which included 1 count and regarded it as the target return signal. The method gave no error pixels as shown in Fig. 6(a). For comparison, the 3D reconstruction with data acquired from the conventional Gaussian fitting method, without using any data optimizing algorithm is also shown in Fig. 6(b). About 11.4% error pixels were registered, most of them lay in the positions of the ears, eyes, nose and hands of the panda where the return signals were too low to be distinguished from the noise counts. As a result, conventional Gaussian fitting could not give the correct data of the range measurement.

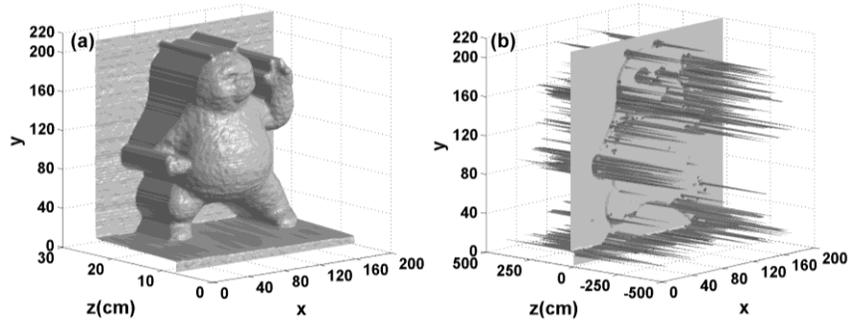

Fig. 6. 200 × 220 pixel depth profile measurements on the PVC toy from a stand-off distance of 2.5 m. (a) Image reconstructed from the raw data using the highest channel of the recorded return histograms. (b) Image reconstructed using the peaks of the Gaussian approximation.

The above results indicated that the simple discrimination method, combined with the low-TJ SNSPD and sub-picosecond time bin, allowed us to obtain the correct depth information for a low signal count target. In addition to the depth information obtained from the TOF, the intensity/counts from the target may also provide interesting information. The plot shown in Fig. 7(b) used colored contour lines to map the calculated numbers of detected return photons for each individual pixel, which could be regarded as the reflectivity imaging of the target object. The color bars directly indicated the logarithm (base 10) of the return photon counts. The reflectivity was related to several properties of the target, such as the surface material, roughness, and the incidence angle, which was hard to distinguish. By combining the depth ranging and reflectivity information, the detailed 3D reconstruction images of the front view (Fig. 7(c)) and lateral view (Fig. 7(d)) were generated using Matlab, with reflectivity data rendered in greyscale. Most of the fine structural features of the target such as the ears, eyes, nose, and fingers were registered.

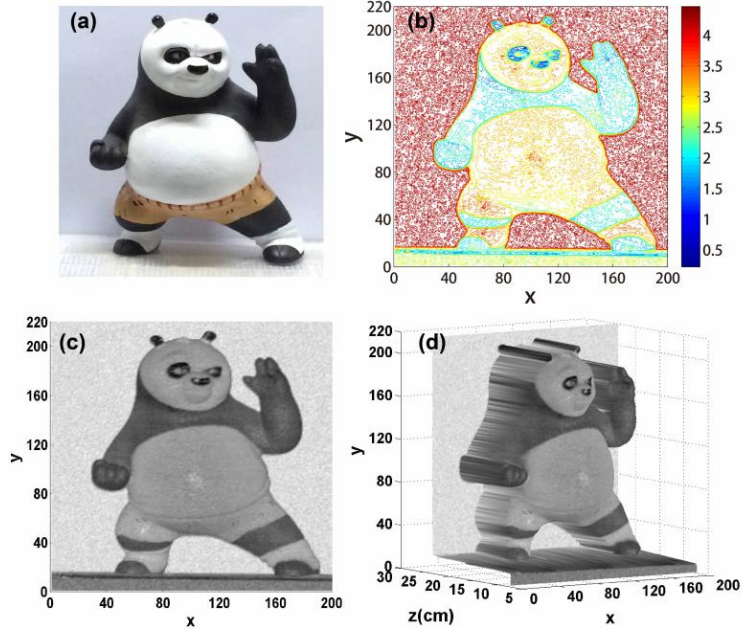

Fig. 7 (a) Close-up photograph of the target object. (b) Reflectivity reconstruction of the target with the common logarithm of the return photon numbers. (c) and (d), 3D reconstructions of the front and lateral views overlaid with reflectivity data.

To further prove the feasibility of this method for few-photon imaging in outdoor noisy environment, the similar experiment was carried out in outdoor daylight condition. The experimental parameters such as the laser power and the stand-off distance were kept same as the indoor experiment. Nevertheless, due to the strong background noise in the daylight condition shown in Fig. 5(b), the total retro-reflected photons including the signal photons and noise photons in some pixels were so high that the SNSPD latched [19, 20]. Thus, we lowered the bias current of SNSPD slightly to 85% of its switching current to prevent the latching problem. The 3D reconstruction images were shown in Fig. 8. Owing to the strong background noise shown in Fig. 5(b), the discrimination threshold was set to 3. About 99.8% of the 200 × 220 pixels exhibited reasonable depth results with the simple discrimination method. The 0.2% error pixels, which could be seen more clearly in lateral view in Fig. 8(b), happened at some locations at the background paper and the low-signature parts of the panda. The error pixels at the background paper were caused by the latching of SNSPD which was not been totally avoided. The SNSPD latched then gave a wrong target signal. The error pixels on the panda were mainly caused by the low retro-reflected signal photon counts, which were comparable to the relative high background noise from the sunshine. These error pixels could be reduced by further reducing the bias current of SNSPD and increasing the laser power.

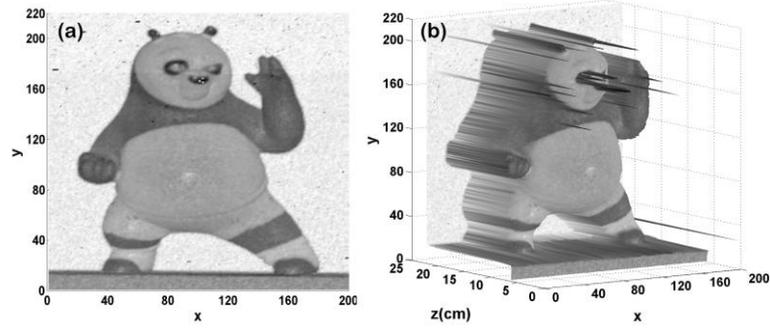

Fig. 8. 3D reconstructions of the front (a) and lateral views (b) overlaid with reflectivity data in outdoor measurement.

## 3. Conclusion

We demonstrated a TCSPC imaging system that was operated at the wavelength of 1550 nm using a low-jitter SNSPD. The 44-ps system TJ, combined with a sub-picosecond time bin, allowed us to easily discriminate the signal counts from the noise signals at a very low return photon level. As a result, laser depth imaging was realized without the use of fitting or algorithms. The fine structural features of the target were registered in the 3D image both indoors and outdoors. This method is interesting for the ranging of low-signature targets and extending the ranging distance. By further decreasing TJ, ranging and imaging with only two retro-reflected signal photons will become possible.

**Acknowledgments**

Hui Zhou and Yuhao He contributed equally to this work. We were grateful for the discussion with G. Wu from the State Key Laboratory of Precision Spectroscopy, East China Normal University. This study was supported by the "Strategic Priority Research Program (B)" of the Chinese Academy of Sciences (Grant XDB04020100&XDB04010200), the National Natural Science Foundation of China (Grant 61401441) and the 973 Program (Grant 2011CBA00202).